\documentclass[%
 aip,
 amsmath,amssymb,
 reprint,%
]{revtex4-2}

\usepackage{xcolor}
\usepackage{graphicx}
\usepackage{dcolumn}
\usepackage{bm}

\usepackage[utf8]{inputenc}
\usepackage[T1]{fontenc}
\usepackage{mathptmx}
\usepackage{etoolbox}
\usepackage{hyperref}

\makeatletter
\def\@email#1#2{%
 \endgroup
 \patchcmd{\titleblock@produce}
  {\frontmatter@RRAPformat}
  {\frontmatter@RRAPformat{\produce@RRAP{*#1\href{mailto:#2}{#2}}}\frontmatter@RRAPformat}
  {}{}
}%
\makeatother

\begin{document}

\preprint{AIP/123-QED}

\title[]{4H-SiC microring opto-mechanical oscillator with a self-injection locked pump}


\author{Anatoliy Savchenkov}
 \affiliation{Jet Propulsion Laboratory, California Institute of Technology, 4800 Oak Grove Drive, Pasadena, California 91109, USA}

\author{Jingwei Li}%
\affiliation{Department of Electrical and Computer Engineering, Carnegie Mellon University, Pittsburgh, Pennsylvania 15213, USA}%

\author{Ruixuan Wang}%
\affiliation{Department of Electrical and Computer Engineering, Carnegie Mellon University, Pittsburgh, Pennsylvania 15213, USA}%

\author{Andrey B. Matsko}
\affiliation{Jet Propulsion Laboratory, California Institute of Technology, 4800 Oak Grove Drive, Pasadena, California 91109, USA}%

\author{Qing Li}%
\affiliation{Department of Electrical and Computer Engineering, Carnegie Mellon University, Pittsburgh, Pennsylvania 15213, USA}%

\author{Hossein Taheri}
 \thanks{Corresponding author: hossein.taheri@ucr.edu}

\affiliation{Department of Electrical and Computer Engineering, University of California Riverside, 3401 Watkins Drive, Riverside, CA 92521}%




\date{\today}

\begin{abstract}
We have demonstrated, for the first time to our knowledge, self-injection locking of a distributed feedback (DFB) diode laser to a multimode 4H-silicon carbide (4H-SiC) microring resonator, and observed resonant opto-mechanical oscillation in the cavity modes. While the fundamental transverse-electric mode family of the silicon carbide microring was optically pumped, Stokes light was generated in the adjacent fundamental transverse-magnetic resonant mode. The threshold of the process did not exceed 5~mW of light entering the cavity characterized with a loaded optical quality factor of $\smash{2\times10^6}$. These results mark a significant milestone in unlocking the potential of 4H-SiC through turnkey soliton microcomb generation and empowering future advancements in areas such as cavity optomechanics using this versatile and quantum-friendly material platform. \copyright 2024. All rights reserved. 
\end{abstract}

\maketitle


\section{\label{sec:intro}Introduction}

\noindent 
Lasers based on resonant opto-mechanical oscillations, including Stimulated Brillouin Scattering (SBS), attract a lot of attention nowadays because of their outstanding spectral characteristics and versatility \cite{Eggleto13aop, alegre2019brillouin, loh15o}. The lasers are used in spectroscopy \cite{domingo05ptl} and quantum optics \cite{Chen22o,Cryer-Jenkins23o} and are optimal for optical clock operation \cite{loh20n}. Because of their high spectral purity \cite{grudinin09prl,Li12oe,gundavarapu19np,Qin22ol}, they can operate over long distances, which is beneficial for telecommunications and local as well as distributed sensing \cite{luan14sr,wang24oe}. Further, they can be integrated into lightweight, compact systems, which is ideal for modern technological applications where weight and space pose constraints.

Resonator host material is critical for operation of an opto-mechanical oscillator. The quality of the emitted light depends on both the mechanical and optical quality factors of the resonator and on its thermal characteristics. Since phase noise of the pump laser can impact the noise of the Stokes light \cite{matsko12oe}, it is important to ensure that the pump light is characterized with low noise. To reduce the noise of the pump light, one may want to lock the pump laser to the nonlinear resonator mode. Furthermore, producing a truly monochromatic emission entails filtering out the pump, which in turn imposes restrictions on the Brillouin frequency offset based on the filter roll-off.

4H-silicon carbide (4H-SiC) is a unique material supporting low-loss mechanical oscillations and a speed of sound approaching 12~km/s in bulk \cite{Vuckovic_4HSiC_nphoton}.  The comparably large frequency offset of the Stokes light generated in crystalline 4H-SiC simplifies optical pump filtering and monochromatic, low-noise Brillouin lasing. 4H-SiC is also a promising platform for scalable, chip-scale, microresonator-based optical frequency comb (microcomb) generation \cite{kippenberg2018ScienceReview, taheri2020review} offering wide bandgap, high thermal conductivity, strong optical Kerr nonlinearity, dispersion engineering by waveguide geometry design, and CMOS compatible microfabrication \cite{Li_4HSiC_octave_comb, taheri2017high-order, Vuckovic_4HSiC_soliton, OuXin_soliton}. Banking on its optically active defect centers and long room-temperature coherence times, 4H-SiC furthermore poses as a desirable material for quantum sensing and hybrid quantum systems \cite{SiC_quantum_review}.

A key enabler of fieldable practical applications of integrated systems in any material platform is turnkey generation and simplified stabilization. Notably, self-injection locking (SIL) \cite{Liang10ol,alkhazraji23lsa} of the frequency of a low-cost, off-the-shelf diode laser to an optical mode of a high-$Q$ (quality factor) microresonator creates a compact optical source addressing simultaneously complexity, size, power consumption, stability, and the reduction of the pump phase noise. We demonstrate here, for the first time to our knowledge, SIL and resonant opto-mechanical oscillation (OMO) in photonic integrated 4H-SiC microring resonators. The ability to achieve OMO with a low threshold of under 5 mW in 4H-SiC microrings with a loaded $Q$ of $\smash{2\times10^6}$ highlights the material's potential for efficient nonlinear interactions, hence opening the door to future advancements in integrated quantum photonics and cavity optomechanics. Importantly, the observed OMO frequency offset is nearly three times smaller compared with the offset value in bulk silicon carbide. The scattering is observed in the forward direction, unlike the backward direction scattering corresponding to SBS. This difference underscores the importance of the geometry of the microphotonic structure and is attributed to both optical and acoustic confinement effects which modify the OMO phase matching conditions \cite{shelby1985brillouin, matsukawa_2004_sbs, rakich2012giant, rakich_2013_tailorable}.

\section{\label{sec:results}Results}

\subsection{4H-SiC microresonators}

The SiC devices were fabricated on a 700-nm-thick 4H-SiC-on-insulator wafer (NGK Insulators) \cite{Li_4HSiC_Raman, Li_4HSiC_soliton}. The pattern, consisting of microrings with both single waveguide and add-drop configurations, was first defined by e-beam lithography (FOx-16 as the e-beam resist) and then transferred to the SiC layer using an optimized CHF$_3$/O$_2$ dry etching process \cite{Li_4HSiC_octave_comb}. In this work, the etch depth was controlled to be near 570 nm, leaving an unetched SiC layer (pedestal) with a thickness of 130 nm. After cleaning, another 2-$\mu$m-thick PECVD oxide layer was deposited to encapsulate the SiC devices. In addition, inverse tapers were implemented near the facet of the SiC by tapering down the SiC waveguide width to approximately 250 nm, which has shown a peak coupling efficiency near $50\%$ (3 dB) to lensed fibers with a mode diameter of $2.5\ \mu$m \cite{Li_4HSiC_soliton}. 

\begin{figure*}[htbp]
    \centering    \includegraphics[width=0.8\textwidth]{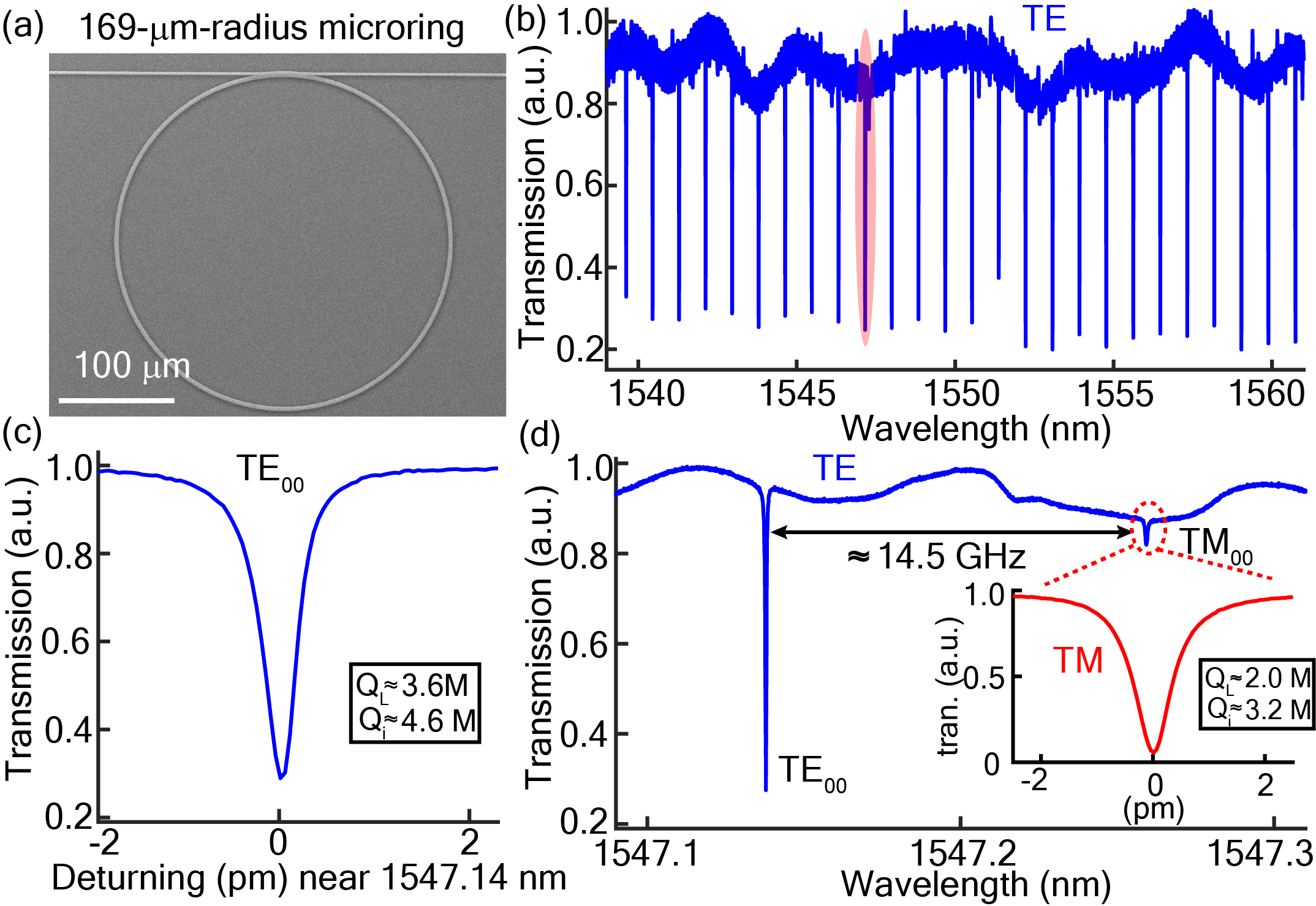}
    \caption{(a) Scanning electron micrograph of a single-side-coupled 169-$\mu$m-radius SiC microring with width of 3 $\mu$m. (b) Linear transmission of the TE-polarized light in the 1550 nm band. The highlighted resonance near the wavelength of 1547 nm is employed for the self-injection locking experiment. (c) A close-up view of the TE$_{00}$ resonance showing a loaded and intrinsic quality factor around $3.6$ million and $4.6$ million, respectively, and (d) examination of the optical spectrum reveals an adjacent TM$_{00}$ resonance near the pump mode with a red-shifted frequency around $14.5$ GHz, which is excited due to residual TM polarization when the input laser is dominated by the TE polarization. The TM$_{00}$ becomes obvious once we change the polarization to TM, which shows a loaded and intrinsic quality factor around $2.0$ and $3.2$ million, respectively.}
    \label{fig:modes}
\end{figure*}

As shown in Fig.~\ref{fig:modes}(a), the microring resonator employed in this work has a radius of 169 $\mu$m, corresponding to a free spectral range (FSR) near 100 GHz \cite{Li_4HSiC_soliton}. The ring width was chosen to be 3 $\mu$m to ensure low propagation loss inside the microring. To selectively couple to the fundamental transverse-electric mode (TE$_{00}$), we adopted a pulley coupling scheme consisting of an access waveguide width of $1.45\ \mu$m and a coupling length of $30\ \mu$m (see Fig.~\ref{fig:modes}(a)). The linear transmission provided in Fig.~\ref{fig:modes}(b) confirms that the TE$_{00}$ mode family is efficiently excited for a coupling gap of 250 nm. As seen in Fig.~\ref{fig:modes}(c), the TE$_{00}$ resonance of this single-side-coupled microring possesses a loaded $Q$-factor of $3.5$ million (resonance bandwidth of $\sim$55 MHz) and an intrinsic $Q$ factor near $4.6$ million. To observe resonant SBS, discussed in Section~\ref{sec:SBS}, the presence of another mode separated from the pumped mode of the resonator by the OMO shift is necessary. It turns out that our pulley coupling scheme can also effectively excite the fundamental transverse-magnetic (TM$_{00}$), which has a slightly different FSR than the TE$_{00}$ resonance (5 GHz difference). As such, we can find the spectral separation between the TE$_{00}$ and TM$_{00}$ modes varying at different azimuthal orders. This feature uniquely benefits the SBS process, as it is relatively straightforward to find mode combinations corresponding to the OMO shift. One such example is shown in Fig.~\ref{fig:modes}(d), where the TM$_{00}$ resonant mode was located at a red-detuned frequency of $14.5$ GHz, which can also be slightly adjusted by varying the temperature. Note that the observed shallow dip of the TM$_{00}$ resonance was only because of the input polarization was chosen to be TE. In fact, once we adjust the input polarization to TM, the same TM$_{00}$ resonance is fully excited, showing close to critical coupling with a loaded quality factor near $3.2$ million (see the inset of Fig.~\ref{fig:modes}(d)). 

\subsection{Experimental setup}

\begin{figure*}[htbp]
    \centering    \includegraphics[width=0.8\textwidth]{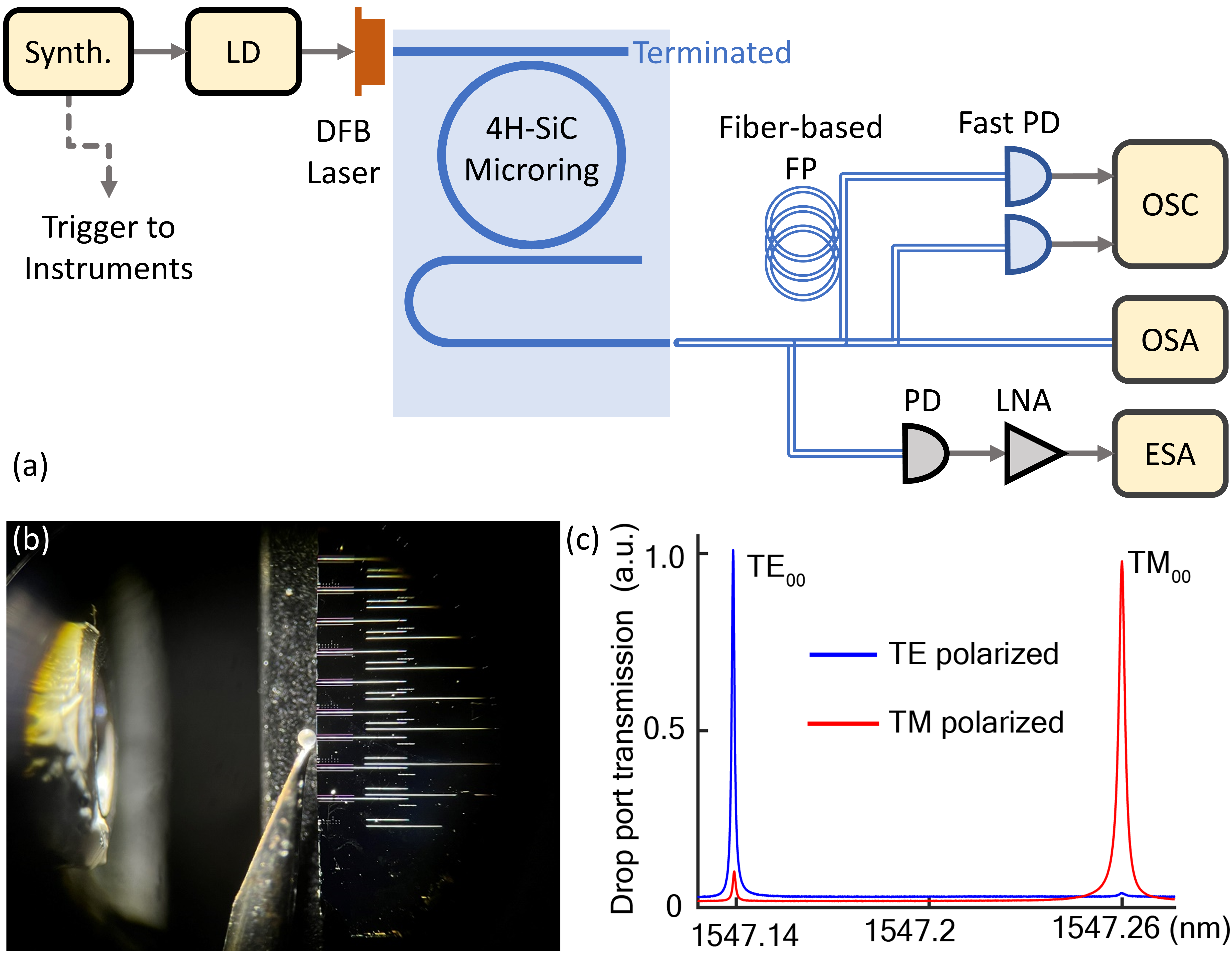}
    \caption{(a) Schematic of the experimental setup. Synth.: Synthesizer, LD: Laser driver, PD: Photodetector, OSC: Oscilloscope, OSA: Optical spectrum analyzer, ESA: Electrical spectrum analyzer, LNA: Low-noise amplifier, FP: Fabry-Perot filter. Waveguide-resonator pulley coupling and waveguide tapers near the chip edges are not emphasized in the schematic. (b) DFB laser coupling to the chip using a ball lens of 300~$\mu\!$m diameter. (c) Transmission of the drop port of the pulley-coupled 169-$\mu$m-radius SiC microring, which has effective coupling to both TE$_{00}$ and TM$_{00}$ mode families. The linewidths of the TE$_{00}$ and TM$_{00}$ resonances shown here are measured to be close to 100 MHz and 210 MHz, respectively.}
    \label{fig:setup}
\end{figure*}

In this work, we mainly employ the single-side-coupled microrings such as the one shown in Fig.~\ref{fig:modes} for intrinsic and coupling quality factor characterization. Their low transmission when on resonance, however, causes problems for experiments such as self-injection locking (SIL). To solve this issue, we implemented add-drop structure as illustrated in the experimental setup in Fig.~\ref{fig:setup}(a), where the transmission is the highest when on resonance. For self-injection locking (SIL), the output light of a distributed feedback (DFB) laser in transistor outline packaging (TO can) at a wavelength of approximately 1546 nm is coupled to the microresonator via a ball lens. A maximum coupling efficiency of 80\% is achieved using a ball lens with a diameter of 300~$\mu\!$m. The emission frequency of the DFB laser is controlled by changing its current via a laser driver. Utilizing a high-power-tolerant DFB TO can, increasing the current to 100 mA and achieving an output optical power of 20-30 mW was possible. The add-drop microring employed in the SIL experiment has the same parameters as the one shown in Fig.~\ref{fig:modes}(a), with the major difference being that another drop port was added to the microring to filter out the useful modes in the transmission measurement. Because of this additional coupling, the linewidth of the TE$_{00}$ resonance has increased to $100$ MHz, corresponding to a loaded $Q$ near $2.0$ million. In addition, we do observe a similar TM$_{00}$ mode family appearing the spectrum (see Fig.~\ref{fig:setup}(d)), which benefits the SBS operation. 

\subsection{Self-injection locking}

\begin{figure*}[htbp]
    \centering    \includegraphics[width=0.9\textwidth]{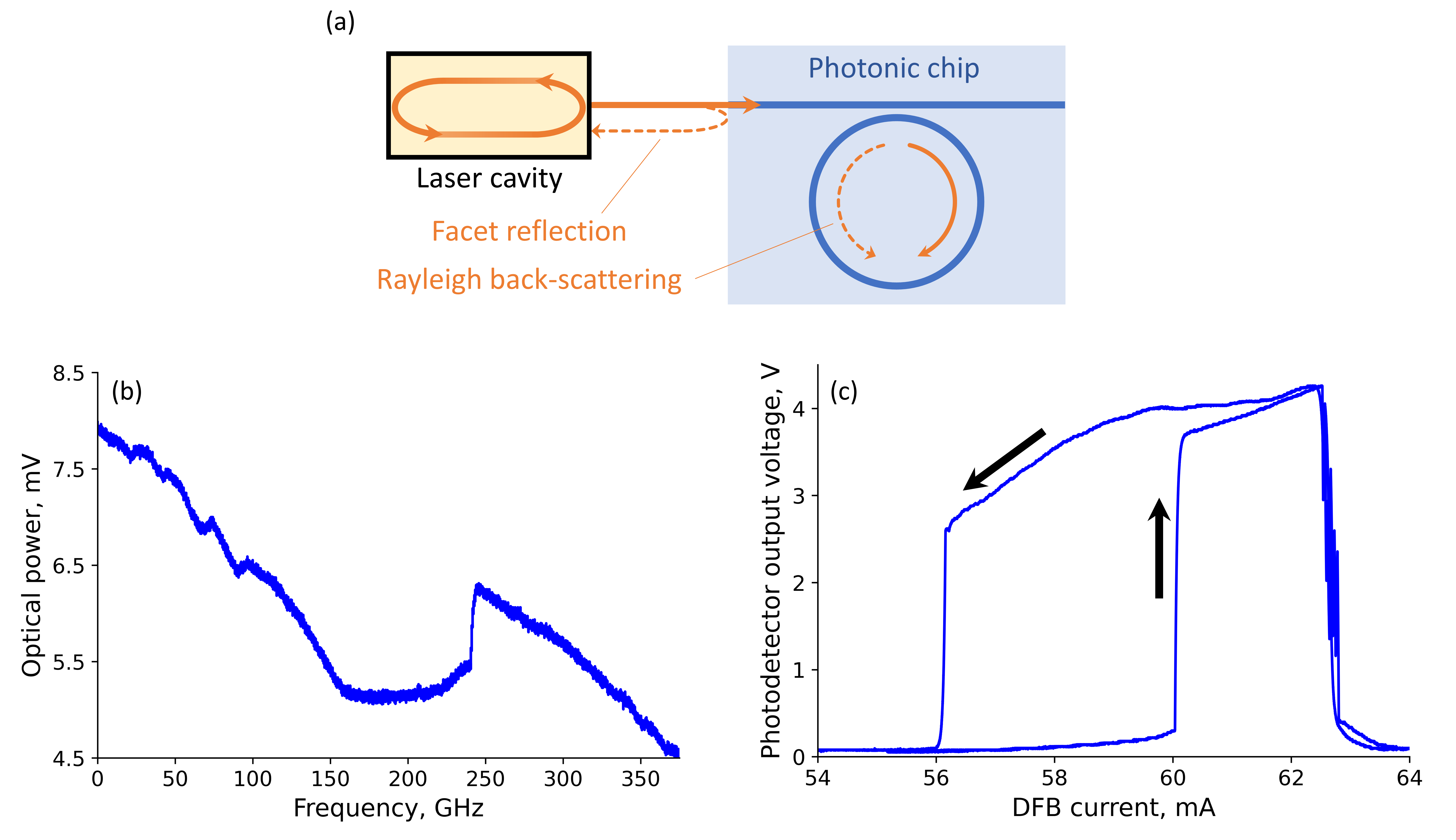}
    \caption{(a) Competition between the reflection from the facet and back-scattering from the microresonator which complicates self-injection locking. The drop port is not shown in this schematicl (b) The self-injection locking (SIL) dip signature in the plot of the detected power versus frequency of the DFB laser as its current is changed. (c) Hysteresis observed in SIL during the forward and backward tuning of the DFB laser current.}
    \label{fig:reflection_SIL}
\end{figure*}

Self-injection locking (SIL) relies on the reflection of the light from the resonator to the laser. It can be facilitated by the Rayleigh back-scattering of light from the optical mode of the microresonator into the laser cavity. It also can be supported by the reflection from the drop coupler through the resonator mode. Laser linewidth after SIL, $\smash{\delta\omega_\mathrm{SIL}}$, can be reduced compared to its free-running value, $\smash{\delta\omega_\mathrm{free}}$ when the light reflected from a high-$Q$ microresonator is efficiently collected. The linewidth reduction $\smash{\delta\omega_\mathrm{free} / \delta\omega_\mathrm{SIL}}$ is proportional with $\smash{Q^2_\mathrm{m} / Q^2_\mathrm{lc}}$, the square of the ratio of the quality factors of the microresonator mode and the laser cavity, and with $\smash{R_\mathrm{m}}$, the relative power reflection from the microresonator mode \cite{kondratiev2017sil}. As a result, the higher the $Q$ of the microresonator and the higher the Rayleigh back-scattering, the better. Achieving SIL in integrated microresonators is particularly challenging because $Q_\mathrm{m}$ is smaller compared with whispering gallery mode resonators, and also because of the competition between the free running frequency of the DFB laser reflected from the chip facet and the Rayleigh back-reflection from the microresonator mode, Fig.~\ref{fig:reflection_SIL}(a). While it is possible to reduced the chip facet reflection by adding an anti-reflection coating on the side of the chip or by an angled facet, we have demonstrated SIL in 4H-SiC microring resonators without taking such measures.

A signature of SIL constitutes a sharp dip in the so-called LI curve (DFB laser power detected by the photodiode versus the laser drive current) \cite{savchenkov2018stiffness}. As the current of the laser is swept, its emission frequency (and power) change monotonically. Because of its relatively large linewidth before SIL, laser emission does not enter the microresonator efficiently. When the laser frequency hits a mode of the resonator, some amount of light reflects back to the laser. The laser locks to the microring mode, the photocurrent (detected power) drops suddenly as a result of the increased power coupling into the microresonator and back-reflection towards the laser cavity. This signature is observed in Fig.~\ref{fig:reflection_SIL}(b). This so called "LI curve" also shows that the competition between the Rayleigh back-scattered and the facet back-reflected light tends to soften the wall of the sharp dip, a feature particularly visible near 150~GHz in Fig.~\ref{fig:reflection_SIL}(b). 

The SIL process corresponds to a three-contour oscillator locking process. The contours include the laser resonator, the microring, and the optical delay line between the laser cavity and the microring. In this configuration the system has several partially overlapping operation branches and there exist two solutions for each value of the current of the SIL laser. As a result, bistability can be observed \cite{kondratiev17oe}. We also observed the phenomenon using the transmission (drop) optical port Fig.~\ref{fig:reflection_SIL}(c). The power of the transmitted light through the resonator was observed as a function of the electric current driving the laser. The transmission was minimal when the light was not locked to the resonator mode. The power jumped up when the laser locked to the microring optical mode. We modulated the laser current with a symmetric saw-tooth signal and observed dissimilar locking ranges for the forward and backward current scan. Such hysteretic behavior is another evidence of the SIL of the laser \cite{kondratiev17oe}. 

\begin{figure*}[htbp]
    \centering    \includegraphics[width=0.8\textwidth]{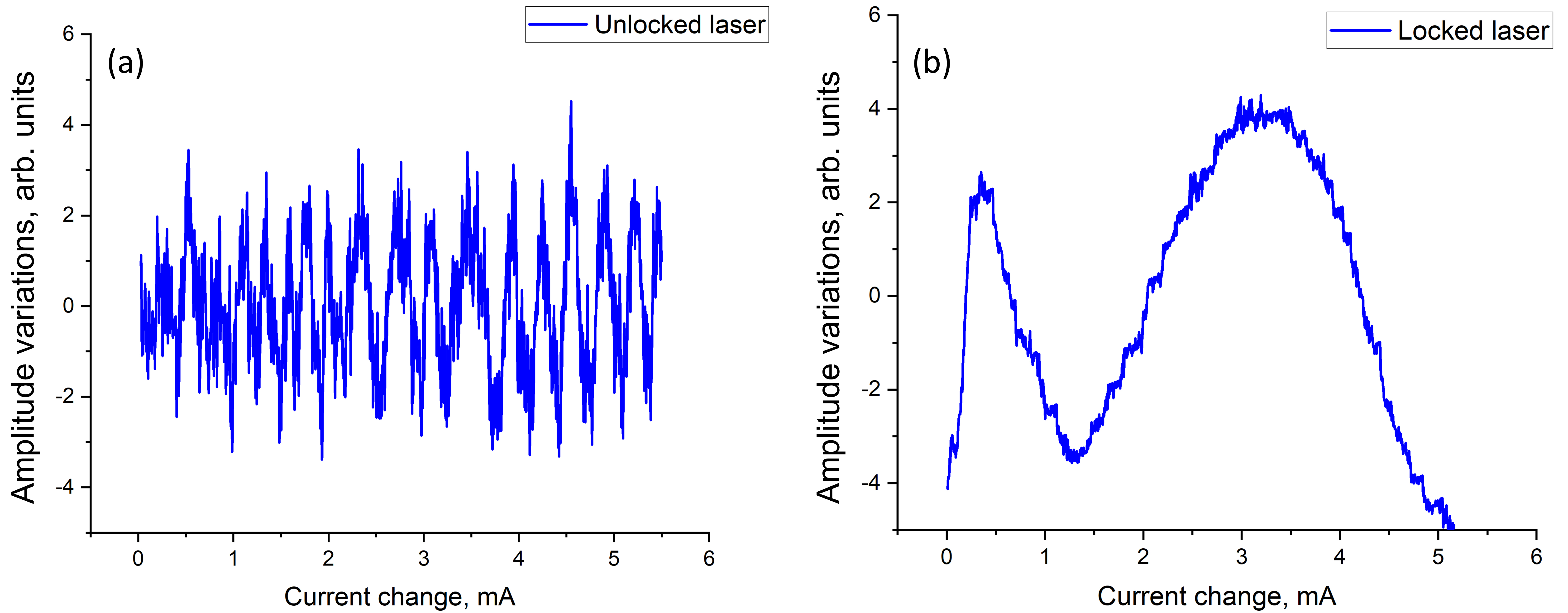}
    \caption{The measured dependence of the laser frequency on the laser current in the (a) unlocked and (b) self-injection-locked laser states. The frequency is measured by sending laser light through an etalon while scanning the laser current. The etalon was a fiber-based low-finesse Fabry-Perot filter; see Fig.~\ref{fig:setup}(a).}
    \label{fig:unlocked_vs_locked}
\end{figure*}

Figure~\ref{fig:unlocked_vs_locked} depicts the photodetector signal as a function of the DFB laser current after passing the optical wave leaving the microresonator through an etalon (a low-finesse fiber-based Fabry-Perot filter) with an FSR of 104~MHz. Without SIL, when the laser emission frequency is scanned by changing its current, this frequency traverses several etalon modes and the photocurrent is hence rapidly modulated, resulting in the oscillatory features observed in Fig.~\ref{fig:unlocked_vs_locked}(a). On the other hand, after achieving SIL, this frequency locks to the nearby microring mode and its tuning rate decreases as the current is changed. As a result the laser emission frequency does not move considerably compared to the etalon modes and the modulation of the photodetector current is significantly reduced; see Fig.~\ref{fig:unlocked_vs_locked}(b).

\subsection{\label{sec:SBS} Observation of resonant optomechanical oscillation}

\begin{figure*}[htbp!]
    \centering    \includegraphics[width=0.8\textwidth]{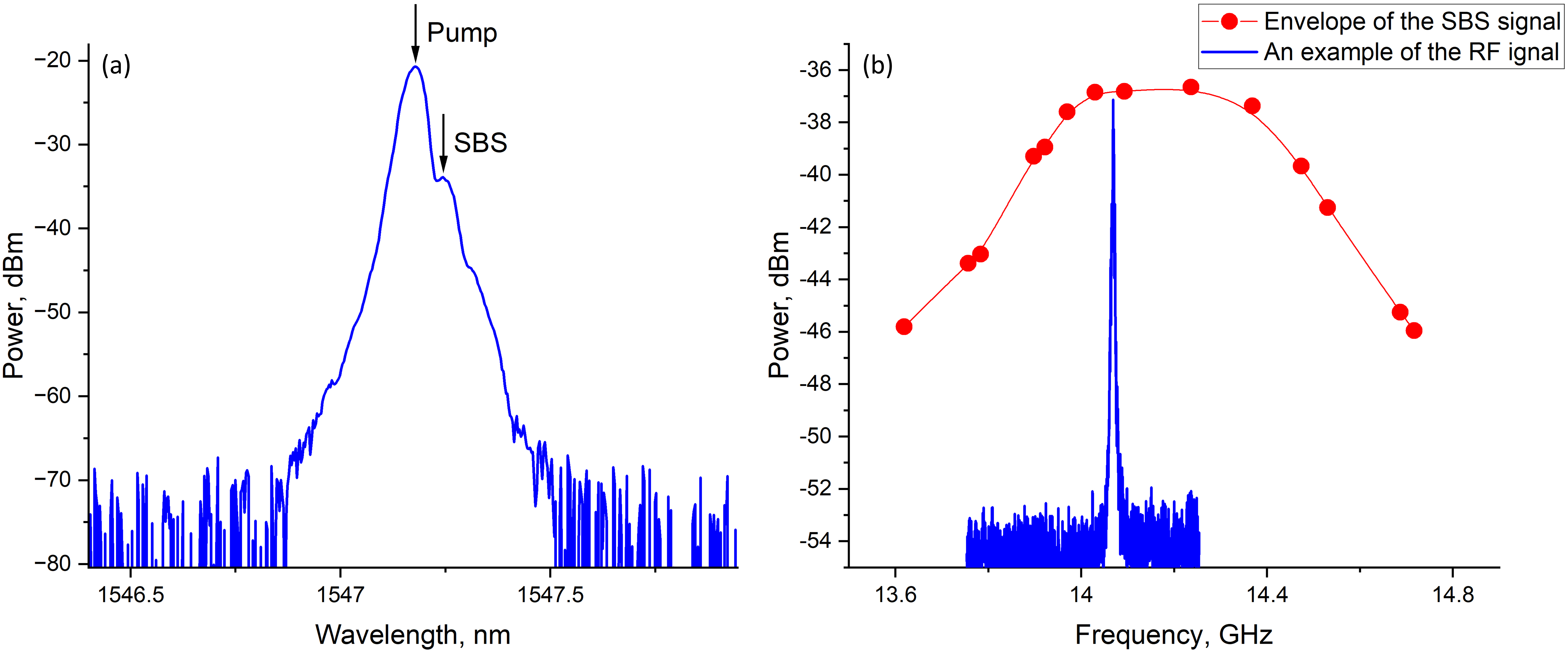}
    \caption{Characterization of the opto-mechanical oscillation. (a) Optical spectrum of the light exiting the resonator shows the presence of the Stokes light. (b) The power recorded by a fast photodiode of the radio-frequency (RF) signal generated by the light with spectrum plotted in (a). The RF signal possesses high spectral purity (blue line). The red dots illustrate the power and peak position of the RF spectral curves for various resonator temperatures.}
    \label{fig:SBS}
\end{figure*}

In this work, we also investigated resonant OMO in 4H-SiC microresonators. Figure~\ref{fig:SBS}(a) illustrates the generation of a Stokes sideband by resonant OMO as laser light is coupled to  the microring. The spectrum of the forward propagating light is recorded from the drop port of the microring; see Fig.~\ref{fig:setup}(a). In this process, the optical pump photons interact with acoustic phonons, leading to the creation of a Stokes signal at a frequency shifted from the original pump by approximately 14.5~GHz. The optical power entering the cavity does not exceed 5~mW and the OMO signal is supported by a mode belonging to a different mode family (the TM mode) than the pumped TE mode. As noted earlier, we verified that the cavity houses a mode at this detuning and can hence support the resonant build up of the OMO sideband. The OMO frequency shift of $14.5$ GHz is defined by the geometry of the system; see Section~\ref{sec:discussion}. This number is also consistent with the linear optical transmission measurement of the microring resonator, seen in Fig.~\ref{fig:setup}(d), revealing that the OMO mode should be cross-polarized compared to the pump (i.e., TM instead of TE). 

The OMO signal in Fig.~\ref{fig:SBS}(a) is partially masked by the pump due to the limited resolution of our optical spectrum analyzer. To ensure the existence of the observed Stokes harmonic, we demodulated the light leaving the resonator on a fast photodiode and characterized the spectrum of the microwave signal using a radio-frequency (RF) spectrum analyzer. An example of the RF spectrum is the blue curve in Fig.~\ref{fig:SBS}(b). The RF signal is well-defined, signifying good signal to noise ratio and high spectral purity. 

The frequency of the OMO signal can be changed by the modification of the spacing between the optical modes. To verify this hypothesis, we recorded the radio-frequency spectrum and changed the temperature of the setup, which effectively alters the frequency spacing between the two adjacent TE$_{00}$ and TM$_{00}$ resonances considering they have different FSRs. As a result of the temperature variation, the radio-frequency signal was observed to shift in frequency. The power and position of the shifted spectra are illustrated by red dots in Fig.~\ref{fig:SBS}(b). The solid line is a guide for the eye illustrating the bandwidth of the phase matching for the process.

\section{\label{sec:discussion}Discussion}
In nanoscale integrated structures, the backward and especially the forward OMO efficiency has been shown to increase significantly compared to that in bulk material \cite{rakich2012giant}. The opto-mechanical scattering offset frequency is defined by the phase matching conditions. In this scattering process photons of the optical pump with frequency $\omega_\mathrm{p}$ scatter into an optical harmonic with a smaller frequency $\omega_\mathrm{S}$ and an acoustic wave with frequency $\Omega$, where $\Omega = \omega_\mathrm{p} - \omega_\mathrm{S}$. The frequency of the scattering is defined by the assumption that the sound has constant speed and that the phase matching conditions are fulfilled
\begin{equation} \label{phasematch}
    {\bf k}_\mathrm{p} = {\bf k}_\mathrm{S}+{\bf k}_\mathrm{\Omega},
\end{equation}
where ${\bf k}_\mathrm{p}$, ${\bf k}_\mathrm{S}$, and ${\bf k}_\mathrm{\Omega}$ are the wave vectors of the pump, Stokes, and the acoustic waves, respectively. The pump and Stokes waves in opto-mechanical scattering in a resonant structure supporting higher order modes can be either nearly counter- or nearly co-propagating. In counter-propagating scattering, the offset frequency approaches that in bulk, while for co-propagating pump and Stokes waves, it can be much smaller than the bulk value of scattering frequency.

Resonant scattering takes place when the optical cavity can accommodate both the pump and the Stokes optical modes. Indeed, in our experiment we have verified that there exists a mode at the frequency offset which can accommodate the Stokes light, Fig.~\ref{fig:setup}(d). Since the frequency splitting between the modes is significantly smaller than the SBS offset for the material, the mode can support the opto-mechanical process corresponding to the forward scattering.

The forward SBS in bulk material is phase matched for zero frequency offset only. The condition changes for a microcavity. Projecting Eq.~(\ref{phasematch}) along the propagation path of the light and sound, we arrive at the equation for the sound wavevector
\begin{equation}
   k_{{\parallel}\mathrm{\Omega}} = k_{{\parallel}\mathrm{p}}-k_{{\parallel}\mathrm{S}},
\end{equation}
where $k_{{\parallel}\mathrm{p}}$ and $k_{{\parallel}\mathrm{S}}$ are the projections of the optical wave numbers. The difference of the wave vectors of two co-propagating optical modes belonging to two different mode families depends on the resonator geometry and can be relatively large \cite{savchenkov2007stoppedlight}. It means that the scattering process can have a large frequency offset. 

The absolute value of the wave vector of the sound wave is given by
\begin{equation}
    |{\bf k}_{\Omega}|^2=\frac{\Omega^2}{v_{\Omega}^2}=k_{\mathrm{cutoff}}^2+k_{\parallel\Omega}^2=k_{\mathrm{cutoff}}^2+(k_{{\parallel}\mathrm{p}}-k_{{\parallel}\mathrm{S}})^2,
\end{equation}
where $k_{\mathrm{cutoff}}$ is defined by the geometry of the system, $\Omega$ is the sound wave frequency, and $v_{\Omega}$ is the speed of sound. When the waveguide cross section is comparable with the wavelength of sound, sound velocity will be small and $|{\bf k}_{\Omega}| > k_{\parallel\Omega}$ can be achieved. In this case, the scattering frequency is influenced by the cut-off frequency of the sound wave. We believe that this effect impacts our observations.

The geometrical confinement also simplifies reaching the  threshold of the opto-mechanical oscillation. Indeed, the threshold power of the process can be estimated as
\begin{equation}
    P_{\mathrm{th}} = \frac{\pi^2 n_\mathrm{p} n_\mathrm{S}}{g_\mathrm{OMO} Q_\mathrm{p} Q_\mathrm{S}} \frac{V}{\lambda_\mathrm{p} \lambda_\mathrm{S}},
\end{equation}
where $n_\mathrm{p}$ ($n_\mathrm{S}$), $Q_\mathrm{p}$ ($Q_\mathrm{S}$), and $\lambda_\mathrm{p}$ ($\lambda_\mathrm{S}$) are the refractive index, quality factor, and wavelength of the pump (Stokes) mode, respectively; $g_\mathrm{OMO}$ is the optomechanical gain defined by the material properties, and $V$ is the mode volume. For the sake of simplicity, we assume that the modes nearly overlap in space. The equation shows that increase of the confinement of the light as well as increase of the quality factors of the modes result in the threshold reduction. Silicon carbide microrings have a great potential for increase of their $Q$-factors, resulting in the reduction of the oscillation threshold. For example, the current intrinsic $Q$ around $4.6$ million is mostly limited by the scattering loss and surface absorption loss, both of which can be improved by optimizing the nanofabrication process in future.  Simultaneously, by optimizing the geometry and modifying the fabrication process through undercutting the structure, the efficiency of the mechanical oscillation can be enhanced.

\section{Conclusion}
We have demonstrated high-frequency opto-mechanical oscillations in a 4H-SiC microring cavity integrated on a chip. The cavity was pumped with a semiconductor laser self-injection locked to a 4H-SiC microresonator mode. No optical amplifier was involved. The observation paves the way towards the realization of silicon carbide turnkey microcomb sources and oscillators on a chip.  

\section*{Funding}
HT and QL were supported by the National Science Foundation of the United States of America under Grant No.~2131402 and 2131162, respectively. Research conducted by AS and ABM was carried out at the Jet
Propulsion Laboratory, California Institute of Technology, under a contract with the National Aeronautics and Space Administration (80NM0018D0004). The CMU team acknowledges the use of Bertucci Nanotechnology Laboratory at Carnegie Mellon University supported by grant BNL-78657879 and the Materials Characterization Facility supported by grant MCF-677785. J.~Li also acknowledges the support of Benjamin Garver Lamme/Westinghouse Graduate Fellowship from CMU. 

\section*{Disclosures} The authors declare no conflicts of interest.

\section*{Data Availability Statement}

The data that support the findings of this study are available from the corresponding author upon reasonable request.

\section*{References}
\bibliography{ref}

\end{document}